\documentclass[apj]{emulateapj}
\usepackage{xspace,graphics,graphicx}
\usepackage{natbib}
\newcommand{\mkms}{{\rm \; km\;s^{-1}}}

\slugcomment{Submitted to ApJL} 

\shorttitle{Emission from galactic outflows}
\shortauthors{Rubin et al.}

\begin{document}

\title{Low-ionization Line Emission from Starburst Galaxies: A New Probe of Galactic-Scale Outflows}
\author{Kate H. R. Rubin\altaffilmark{1,2}, J. Xavier Prochaska\altaffilmark{1}, Brice M{\'e}nard\altaffilmark{3}, Norman Murray\altaffilmark{3}, Daniel Kasen\altaffilmark{1}, David C. Koo\altaffilmark{1} \& Andrew C. Phillips\altaffilmark{1}}
\altaffiltext{1}{University of California Observatories, University of California, Santa Cruz, CA 95064}
\altaffiltext{2}{rubin@ucolick.org}
\altaffiltext{3}{Canadian Institute for Theoretical Astrophysics, 60 St.\ George St, University of Toronto, Toronto, ON M5S 3H8, Canada}

\begin{abstract}
We study the kinematically narrow, low-ionization line emission from a bright, starburst galaxy at $z = 0.69$ using slit spectroscopy obtained with Keck/LRIS.  
The spectrum reveals strong absorption in \ion{Mg}{2} and \ion{Fe}{2} resonance transitions with Doppler shifts of $-200$ to $-300 \mkms$, indicating a cool gas outflow.  
Emission in \ion{Mg}{2} near and redward of systemic velocity, in concert with the observed absorption, yields a P Cygni-like line profile similar to those observed in the Ly$\alpha$ 
transition in Lyman Break Galaxies.  Further, the \ion{Mg}{2} emission is spatially resolved, and extends significantly beyond the emission from stars and \ion{H}{2} regions within the galaxy.  
Assuming the  
emission has a simple, symmetric surface brightness profile, 
we find that the gas 
extends  
to distances $\gtrsim 7$ kpc.  We also detect several narrow \ion{Fe}{2}* fine-structure lines in emission near the systemic velocity, arising from energy levels which are radiatively excited directly from the ground state.  We suggest that the \ion{Mg}{2} and \ion{Fe}{2}* emission is generated by photon scattering in the observed outflow, and emphasize that this emission is a generic prediction of outflows.  
These observations provide the first direct constraints on the minimum spatial extent and morphology of the wind from a distant galaxy.  Estimates of these parameters are crucial for understanding the impact of outflows in driving galaxy evolution.

\end{abstract}
\keywords{galaxies: ISM --- galaxies: halos --- galaxies: starburst}

\section{Introduction}\label{sec.intro}
	
Galactic-scale outflows may be a primary driver of galaxy evolution through the removal of cool gas from star-forming regions to a galaxy's halo or beyond.
Measurements of outflow properties in distant galaxies, such as the mass and energy outflow rates, are central to understanding their role in feedback and quenching processes.  However, accurate determinations of these outflow rates require knowledge of the spatial extent of the wind, a parameter that is difficult to constrain observationally.  

While winds are detected in optical line and X-ray emission around local starburst galaxies to distances of several kiloparsecs \citep[e.g.,][]{Heckman1990,Martin1999}, 
because this emission is faint,  
outflows in distant galaxies 
have traditionally been detected only in absorption against the stellar continuum.  This technique probes cool gas ($T \lesssim 10^4$ K) traced by transitions such as \ion{Na}{1} D \citep[e.g.,][]{Rupke2005a} and \ion{Mg}{2} \citep[e.g.,][]{Weiner2009,RubinTKRS2009}, and constrains the outflow velocity, covering fraction, and column density.  Recently, \citet{Menard2009_Mg2SFH} also showed that \ion{Mg}{2} absorption tracks the overall star formation history of the Universe.  However, these absorption-line studies provide only weak constraints on the radial extent, spatially-resolved geometry, and volume density of the outflow.  
In turn, estimates of the rates of mass and energy loss
are uncertain by at least two orders of magnitude. 
Improved constraints are critical for understanding the role played by outflows in polluting a galaxy's halo and the surrounding intergalactic medium.  

A novel technique for studying outflows in the distant universe is analysis of emission from the outflowing gas.  This emission may arise from resonance-line scattering off of the flow or fluorescent radiation powered by luminous star clusters.  In the former case, the emission, with the corresponding absorption, may exhibit a P Cygni-like line profile.  The high optical depth of the Ly$\alpha$ transition results in P Cygni profiles observed in the spectra of
Lyman Break Galaxies \citep[LBGs; e.g.,][]{Pettini2001,Pettini2002}. 
Although it has much lower optical depth than Ly$\alpha$, 
\ion{Mg}{2} $\lambda \lambda 2796, 2803$ photons can also be resonantly trapped and will produce strong emission where the gas is optically thick (i.e., at hydrogen column densities of $N_H \sim 10^{19}~\rm cm^{-2}$).  
In contrast, optically thick \ion{Fe}{2} resonance absorption is not trapped, but pumps UV \ion{Fe}{2}* emission lines at $\sim$ 2000 - 3000 \AA.
In principle, \ion{Mg}{2} and \ion{Fe}{2}* emission can be particularly useful for tracing winds at $z > 0.3$, where it is detected into the optical; moreover, \ion{Mg}{2} emission has already been detected in $z \sim 1$ star-forming galaxies \citep{Weiner2009,RubinTKRS2009}. Measurement of the spatial extent of this line emission provides a firm lower limit on the radial extent of the outflow and constrains the wind morphology.  

In this paper, we examine the line emission from a bright starburst at $z = 0.694$ known to exhibit a P Cygni profile in \ion{Mg}{2} and \ion{Fe}{2}* emission \citep{Rubin2009}.  
We report in detail on the observed characteristics of the emission and suggest possible production mechanisms.       
We adopt a $\Lambda$CDM cosmology with $H_0 = 70~\rm km~s^{-1}~Mpc^{-1}$, $\rm \Omega_{M} = 0.3$, and $\rm \Omega_{\Lambda} = 0.7$.

\section{Observations}\label{sec.observations}

Details of our observations of the target galaxy (TKRS4389) and data reduction are given in \citet{Rubin2009}.  
We obtained spectroscopy of this galaxy using the Low Resolution Imaging Spectrometer (LRIS) on Keck 1 \citep{Cohen1994}.  Our instrumental setup afforded a FWHM resolution ranging between $190- 300~\rm km~s^{-1}$ and wavelength coverage of $\sim 3200 - 7600$~\AA.  We used a $0.9\arcsec$ slitlet oriented NE 
\citep[see Figure 1 of][]{Rubin2009} and collected six $\sim 1800\rm~ sec$ exposures with FWHM $\sim 0.6\arcsec$ seeing.  
The data were reduced using the XIDL LowRedux\footnote{http://www.ucolick.org/$\sim$xavier/LowRedux/} data reduction pipeline.

\begin{figure}
\begin{center}
\includegraphics[angle=90,width=\columnwidth]{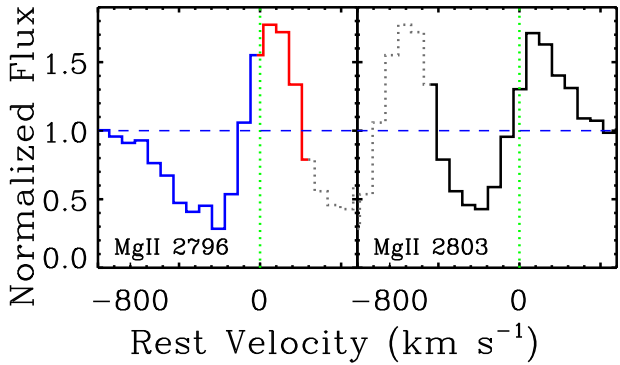}
\caption[MgII P Cygni profile.]{The \ion{Mg}{2} line profile in the galaxy spectrum.  Dotted portions of the 
spectrum 
are associated with the transition in the opposite panel.  
The systemic velocity 
is marked 
with vertical dotted lines, and the horizontal dashed line marks the continuum level.  The profile exhibits 
blueshifted absorption extending to $\rm -800~km~s^{-1}$ relative to the systemic velocity.  
Emission at and redward of systemic velocity is also evident.  Together,  
these features 
exhibit a P Cygni profile, suggestive of a galactic outflow.  
We propose that the red and blue sections of the spectrum in the left-hand panel arise from different areas of the outflow, as indicated in Figure~\ref{fig.cartoon}.
\label{fig.PCygni}}
\end{center}
\end{figure}

\begin{figure}[ht]
\begin{center}
\includegraphics[scale=0.8,angle=90,viewport=0 0 400 300,clip]{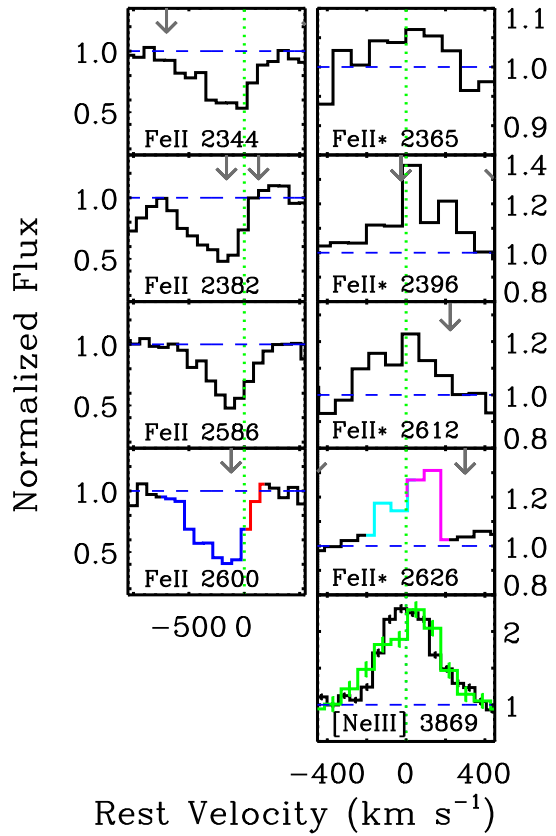}
\caption{ \ion{Fe}{2} transitions in the galaxy spectrum.  The systemic velocity is marked 
with vertical dotted lines, and the horizontal dashed line marks the continuum level.  
The left-hand column shows resonance absorption lines, while the 
right-hand column shows \ion{Fe}{2}* emission profiles (top four panels) and the [\ion{Ne}{3}] $\lambda 3869$ line (bottom panel; black).  
The green line in the bottom panel shows the coadd of the detected (and unblended) \ion{Fe}{2}* emission lines.
Gray arrows mark \ion{Fe}{2}* transitions 
arising from states that cannot be radiatively excited directly from the ground state (see \S\ref{sec.discussion} and Table 1). 
We propose that the red and blue sections of the spectrum in the left-hand panel and the magenta and cyan sections in the right-hand panel arise from different areas of the outflow, as indicated in Figure~\ref{fig.cartoon}.
 \label{fig.feii}}
\end{center}
\end{figure}

\begin{figure*}[ht]
\begin{center}
\includegraphics[scale=0.35,angle=90,viewport=-20 0 600 650,clip]{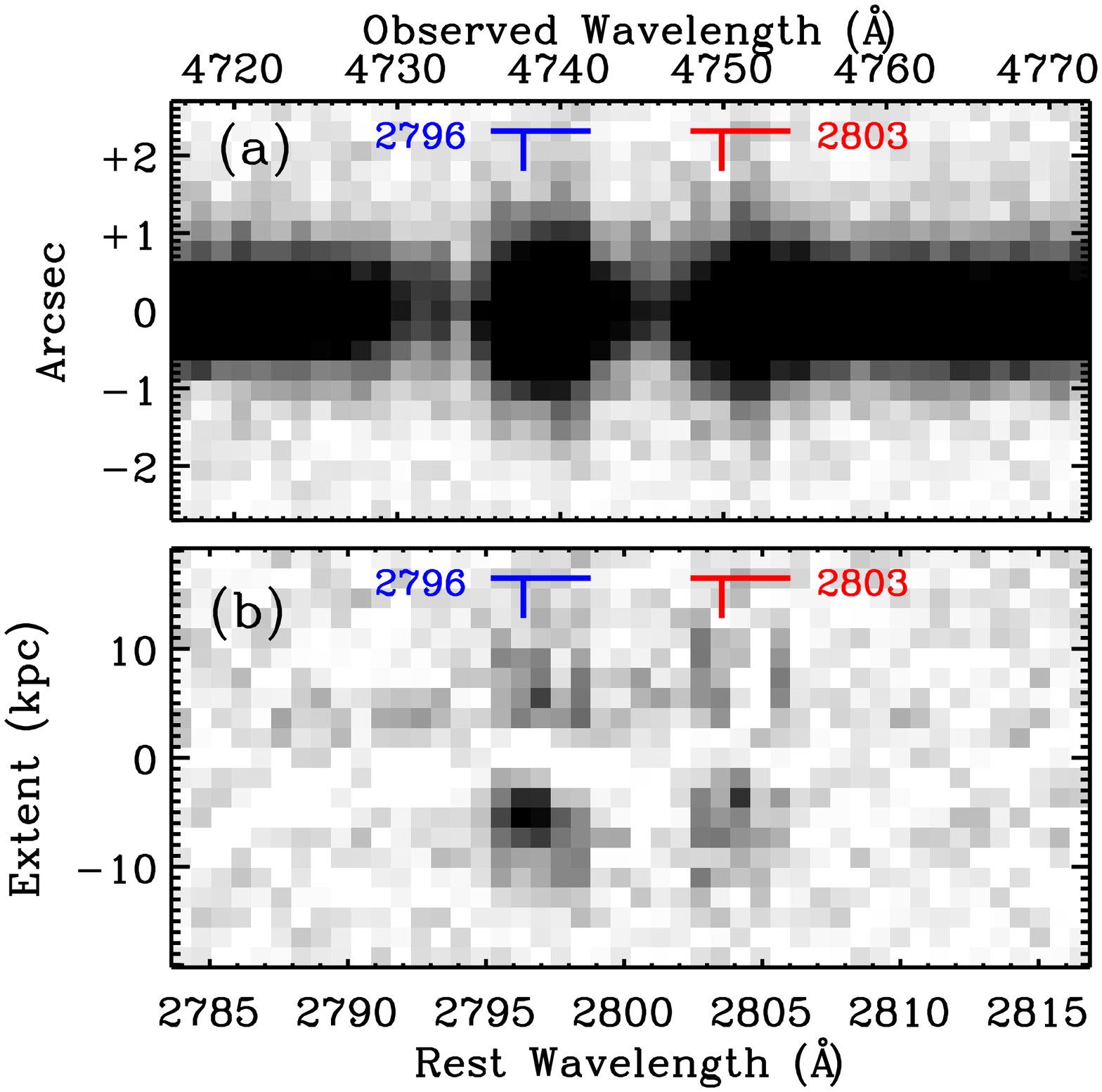}
\includegraphics[scale=0.35,angle=90]{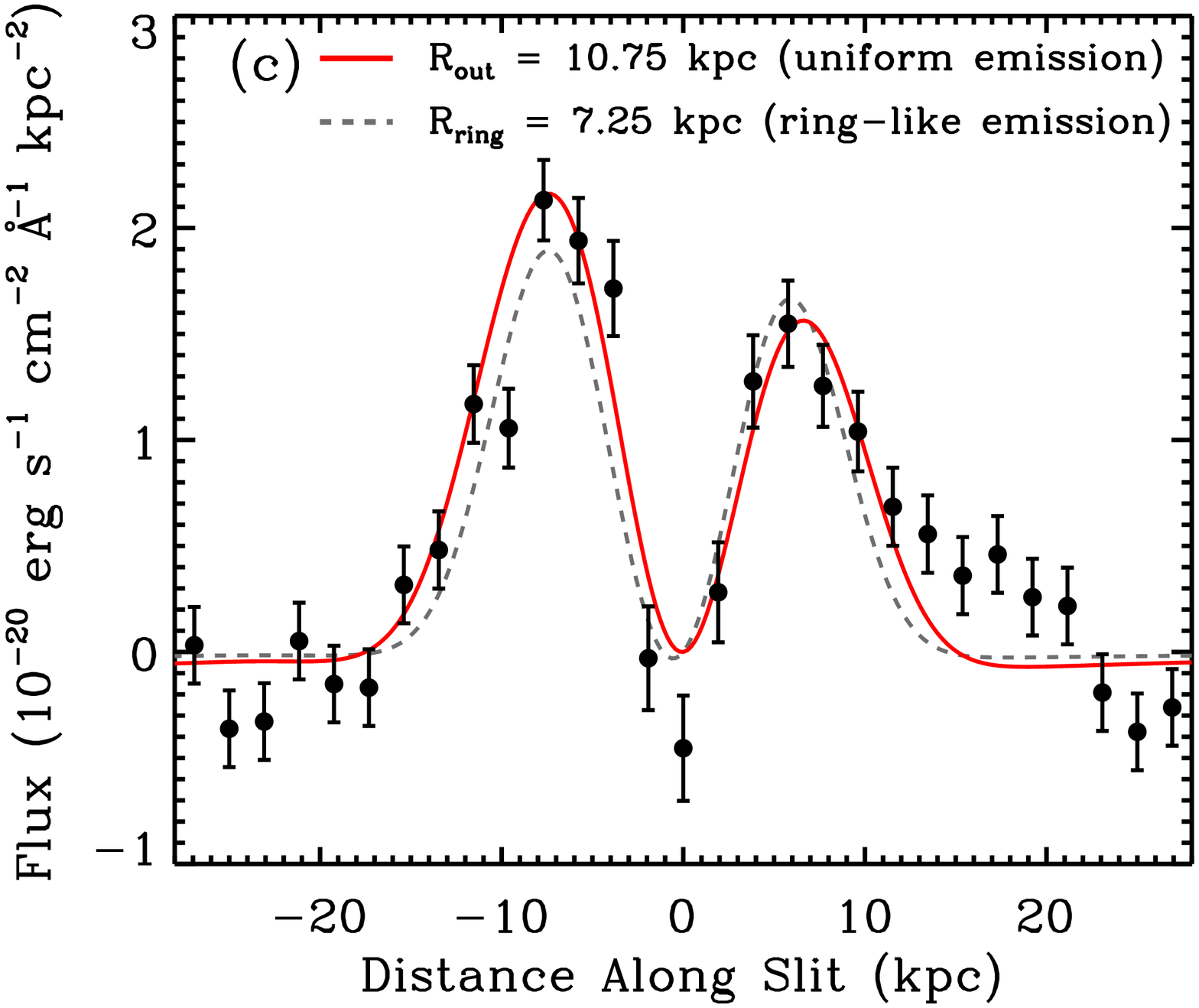}
\caption[Spatially extended MgII emission.]{
{\it (a):} Two-dimensional Keck/LRIS spectrum of the \ion{Mg}{2} doublet, evident in emission and absorption.  The vertical blue and red lines mark the systemic velocities of the \ion{Mg}{2} doublet transitions. {\it (b):} 
A model of the galaxy continuum, scaled by the flux in the spatial center of the object, has been subtracted from the spectrum shown in panel {\it(a)}.  
The residual flux at $\lambda = 2796$ \AA \ and 
$\lambda = 2803$ \AA \ indicates that the \ion{Mg}{2} emission is broader than 
the continuum, and extends on both sides to $\sim 1$\arcsec \ ($\gtrsim 7$ 
kpc).  {\it (c):}  The spatial profile of the spectrum shown in panel {\it (b)}, averaged over the full velocity extent of both emission lines (black circles).  
The lines show the continuum-subtracted emission profile from 
models with different distributions of line-emitting material.  See Section \ref{sec.mg2analysis} for details.

\label{fig.profiles}}
\end{center}
\end{figure*}

\section{Analysis}\label{sec.analysis}

As noted in \citet{Rubin2009}, this galaxy is exceptionally bright for its redshift, with a 
star formation rate of $\sim 80~M_{\odot}~\rm yr^{-1}$.  
Stellar population modeling indicates that the spectrum is dominated by light from 
the intense star formation activity.  
The deepest parts of the \ion{Mg}{2} and \ion{Fe}{2} resonance {\it absorption} 
lines 
are blueshifted by $\sim 200 - 300~\rm km~s^{-1}$, with high velocity tails extending to $-800 \mkms$, 
indicating that these ions 
trace an outflow (see Figures~\ref{fig.PCygni} and \ref{fig.feii}).  Here we analyze the characteristics of 
the observed {\it emission} from these ions.

\subsection{Characteristics of \ion{Mg}{2} Emission}\label{sec.mg2analysis}
The one-dimensional spectrum of this galaxy reveals strong \ion{Mg}{2} emission near systemic velocity and 
extending to the red (Figure~\ref{fig.PCygni}).  
We measure
the kinematic extent $v_{extent}$ of each emission line (Table~\ref{tab.em}) using standard techniques \citep{Cooksey2008}.  
Line fluxes are derived from integrating the continuum-subtracted spectrum over this velocity range.  
The continuum is measured from clean regions near each feature;  its
error is determined through Monte Carlo realizations. 
The flux-weighted velocity centroid of the lines is given as $v_{cen}$, and 
$v_{peak}$ is the brightest pixel in each line (smoothed by 3 pixels; Table 1).

\ion{Mg}{2} absorption and emission are also evident in the two-dimensional spectrum of the target (Figure~\ref{fig.profiles}).  
To explore the spatial extent of the \ion{Mg}{2} emission, we subtract a model of the two-dimensional continuum profile from the original two-dimensional spectrum.  The residual emission extends up to $\sim 2$\arcsec \ from the spatial center of the galaxy continuum, and extends well beyond the stellar structure (i.e., the components of the galaxy which generate the continuum flux near \ion{Mg}{2}).  
The flux of the two emission features at 2796 \AA \ below and 
above the galaxy continuum are $(8.0 \pm  0.4)$ and $(4.4 \pm 0.4) \times 10^{-18}~\rm ergs~cm^{-2}~s^{-1}$, respectively, and have flux-weighted velocity centroids of $\sim 26\mkms$ and $74\mkms$.  The corresponding fluxes at 2803 \AA \ are $(4.0 \pm 0.3)$ and $(2.5 \pm 0.4) \times 10^{-18}~\rm ergs~cm^{-2}~s^{-1}$.  There is a 
significant difference 
in the flux of the 2796 \AA \ line in these two locations;  
this may reflect the geometrical distribution of the emitting gas, suggesting a lack of spherical symmetry. 
Combined, these features contribute most ($\sim 80$\%) of the total 2796 \AA \ 
flux.
In Figure~\ref{fig.profiles}c, we also present the continuum-subtracted spatial profile averaged over the velocity range covered by \emph{both} lines. 
Errors calculated from the extraction of continuum-subtracted spatial profiles at other locations along the slit are consistent with the error bars in the Figure in the wings of the profile and are larger by up to a factor of $\sim 2$ near the profile center.  
Spatially extended emission from strong lines arising in \ion{H}{2} regions, e.g., H$\gamma$ and [\ion{O}{2}], is not detected.

Finally, we develop a rudimentary model for the spatial distribution of the line-emitting material to constrain its distance from the galaxy's center.  
Because \ion{Mg}{2} is emitted in star-forming recombination regions \citep{Kinney1993}, 
the model assumes that the emission is distributed over both the galactic \ion{H}{2} regions and a more extended area with variable geometry and size.
The nebular emission component is distributed in roughly the same way as the young stars in the galaxy; i.e., as in the HST/ACS $b_{435}$-band image.  We 
create a ``continuum" model by scaling this image such that when it is convolved with a two-dimensional Gaussian to simulate the effects of seeing and is ``observed" with the correct slit orientation, 
the spatial distribution of the flux passing through the slit matches the galaxy continuum profile. 
We then assume that 
the extended emission is distributed either (1) with uniform surface brightness and an outer radius $R_\mathrm{out}$, or (2) in a ring with radius $R_\mathrm{ring}$ and thickness $\sim 1$ kpc.  

We generate several realizations of these models by allowing the ratio of the total luminosity from the extended emission to the luminosity in the section of the galaxy centered on the slit ($f_\mathrm{gal}$) to vary between 0.5 and 3 in increments of 0.25, and by varying either $R_\mathrm{out}$ or $R_\mathrm{ring}$ between 5 and 13 kpc in increments of 0.25 kpc.  
These realizations are added to the $b_{435}$ image, and the resulting ``extended emission image" is convolved with a Gaussian and ``observed" through the slit.  We then subtract the profile of the continuum model from the extended emission model profile, first scaling the continuum profile to the peak flux value in the extended emission profile.  
In general, we find that the uniform surface brightness model provides acceptable fits to the data if $1.25 < f_\mathrm{gal} < 2.5$, although adjustments in the value of $R_\mathrm{out}$ are required.  With $f_\mathrm{gal} = 1.5$, $R_\mathrm{out}$ must exceed 9.5 kpc to match the observations.  As $f_\mathrm{gal}$ increases, values in the range 8.25 kpc $ < R_\mathrm{out} < 13$ kpc provide an excellent match to the observed emission (Figure~\ref{fig.profiles}c; red solid line).  
Additionally, though the best-fit uniform surface brightness model yields a lower $\chi^2$ value than any of the ring-like emission models, 
realizations of the latter with 6.5 kpc $\le R_\mathrm{ring} < 8.25 $ kpc and $f_\mathrm{gal} \sim 1.25$ are allowed by the data (gray dashed line).  
While this framework is quite simplistic, the models suggest that the material giving rise to the \ion{Mg}{2} emission 
is distributed over a large area, extending 
to distances of at least 6.5 kpc.

\begin{figure}[ht]
\begin{center}
\includegraphics[width=\columnwidth]{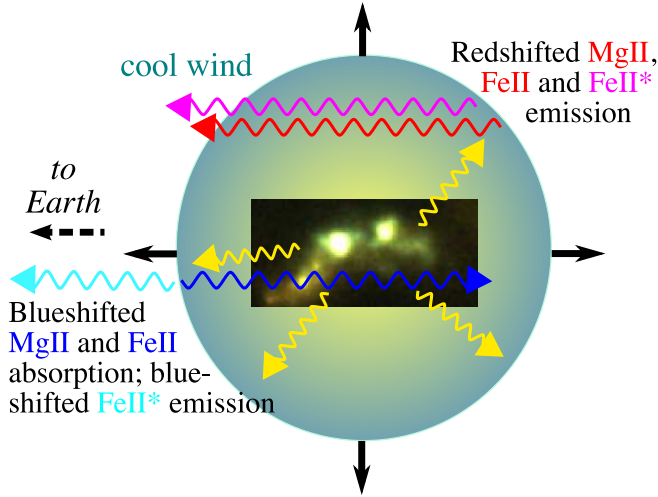}
\caption[Illustration of resonant scattering and fluorescence.]{Cartoon illustrating resonant scattering in \ion{Mg}{2} and fluorescence in \ion{Fe}{2}.  
The galaxy is surrounded by an expanding, spherically symmetric wind.  \ion{Mg}{2} and \ion{Fe}{2} ions in the section of the wind closest to Earth absorb continuum photons at the 
rest frequency of the ions; this produces a blueshifted absorption line profile (shown in blue in Figures~\ref{fig.PCygni} and \ref{fig.feii}).  Ions 
in the back side of the wind scatter photons with wavelengths redshifted relative to systemic velocity; 
these photons travel through the wind and into our line of sight (shown in red in Figures~\ref{fig.PCygni} and \ref{fig.feii}).  Because there are several energy levels accessible to  
\ion{Fe}{2}, the photons are not resonantly trapped, and escape from the wind via emission to excited states (indicated in cyan and magenta in Figure~\ref{fig.feii}). \label{fig.cartoon}}
\end{center}
\end{figure}

\subsection{Characteristics of \ion{Fe}{2}* Emission}\label{sec.id}

The \ion{Fe}{2} ion is notable for its multitude of permitted transitions in the wavelength range 2000 - 3000 \AA\ (Table~\ref{tab.em}).  We observe several \ion{Fe}{2} resonance lines in absorption (Figure~\ref{fig.feii}) which are blueshifted by $\sim -200 \mkms$, similar to \ion{Mg}{2}.  We also identify several transitions in emission (Figure~\ref{fig.feii}).
All of these are consistent with transitions arising exclusively from the $\rm J = 9/2$ or 7/2 upper levels (see Table~\ref{tab.em} for details). 

The velocity limits of each emission line, $v_{cen}$, and $v_{peak}$ are determined as for \ion{Mg}{2}.  Both $v_{peak}$ and $v_{cen}$ are redward or within $30 \mkms$ of systemic in 4 of 5 detected lines.  In an inverse variance-weighted stack of the emission lines at 2365 \AA, 2396.35 \AA, 2612 \AA, and 2626 \AA \ (overplotted in green
in Figure~\ref{fig.feii}), $v_{peak} = +50 \mkms$ and $v_{cen} = +8 \mkms$.
We note that this differs from the velocity profiles of nebular emission lines such as [\ion{Ne}{3}], H$\delta$, and H$\gamma$, which have peak velocities near or blueward of systemic and blueward of $v_{cen}$. 
The [\ion{Ne}{3}] velocity profile is shown in Figure~\ref{fig.feii} as a reference; this line is chosen because the profile is not contaminated by broad stellar absorption.  
The velocity profiles of the \ion{Fe}{2}* emission lines significantly differ both from those of the \ion{Fe}{2} resonance absorption lines and from the profiles of lines tracing the kinematics of the galactic \ion{H}{2} regions. 


\section{Discussion}\label{sec.discussion}
We now suggest a possible production mechanism for the observed \ion{Mg}{2} and \ion{Fe}{2}* emission: 
photon scattering in a large-scale galactic outflow.
As discussed in Section~\ref{sec.intro}, this has been invoked in previous work to explain the P Cygni profile of Ly$\alpha$ emission
in LBGs. 
Similarly to Ly$\alpha$, \ion{Mg}{2} is likely to have high optical depths in the ISM of star-forming galaxies.  
Additionally, the \ion{Mg}{2} transitions of interest arise from a set of two close upper energy levels that 
can only decay to 
the ground level.  \ion{Mg}{2} photons are therefore resonantly trapped in 
high optical depth conditions 
(see also Prochaska et al., in prep).

Figure~\ref{fig.cartoon} illustrates the effects of a transition resonantly trapped in 
an outflow.  \ion{Mg}{2} ions in the section of the wind closest to Earth absorb continuum photons at the \ion{Mg}{2} transition in the rest frame of the gas; this produces a blueshifted absorption line profile.  \ion{Mg}{2} ions 
in the far section of the wind scatter photons with wavelengths redshifted relative to the rest frame of the front part of the wind; these photons 
can therefore travel through the wind, generating emission at and redward of the systemic velocity.  
Such an outflow may also 
produce the spatially extended emission evident in Figure~\ref{fig.profiles}.  This emission arises from 
gas above and below the galaxy in Figure~\ref{fig.cartoon} moving tangentially to our line of sight, and so is expected to occur near the systemic velocity.

A similar mechanism can produce the observed \ion{Fe}{2}* emission.  
Ions in the front side of the wind will again produce blueshifted absorption in resonance transitions, as in the 
case of \ion{Mg}{2}.  The excited ions in {\it all parts} of the wind will then emit photons and decay to states with $\rm |\Delta J| \leq 1$, including \ion{Fe}{2}* transitions to fine-structure levels with J $\ge 5/2$.  
We note that these are the only transitions observed; i.e., we do not detect emission 
lines with lower energy levels having $\rm J = 3/2$ or 1/2, and the detection of these lines 
in higher-S/N data would rule out this picture.  
Further, the observed emission arises from the same upper levels excited in the resonance transitions.  
The lack of observed absorption from these excited states indicates that they decay before the ion is re-excited.  The wind is therefore completely optically thin 
to the observed \ion{Fe}{2}* photons.  

\section{Implications}\label{sec.implications}
The mechanism described above is only one of several which may produce \ion{Mg}{2} and \ion{Fe}{2}* emission.  These lines may be produced in recombination regions \citep{Kinney1993} or in AGN \citep{Vestergaard2001}.  However, because the kinematics of all the lines discussed are qualitatively different from the kinematics of the nebular emission in the spectrum, and because our analysis of the spatial distribution of the \ion{Mg}{2} emission suggests that it arises from distances out to at least $\sim 7$ kpc from the galaxy center, we consider photon scattering the most probable production mechanism for the observed emission.  As such, \emph{our observation represents the first detection of a galactic outflow in emission in a distant galaxy, and provides the first constraints on the minimum spatial extent ($\gtrsim 7$ kpc) and morphology of this outflow}.  Stringent constraints on the mass and energetics of the gas require a solution to the radiative transfer equation in the wind.  However, here we 
use our EW measurements (Table~\ref{tab.em}) to make a crude estimate of the physical properties of the outflow.  The ratio of the \ion{Mg}{2} absorption line EWs (i.e., the ``doublet ratio") is $\sim 1.64$; this indicates that the optical depth is $\tau_{2796} \gtrsim 1.4$ along the line of sight to the galaxy itself \citep{Spitzer1968,Jenkins1986,RubinTKRS2009}.  This analysis does not account for the effects of the emission on the absorption EWs; however, because the 2803 \AA \ absorption line is filled in by emission from both transitions, the EW is likely reduced more than the 2796 \AA \ absorption EW.  We therefore consider this estimate a conservative lower limit.  Assuming a covering fraction equal to 1 and invoking Eq. 2-41 of \citet{Spitzer1968}, $\tau_{2796} \gtrsim 1.4$ implies a column density of $N_\mathrm{MgII} > 10^{14}~\mathrm{cm^{-2}}$.
In the case of an isotropic wind which is fully covered by the slit, the sum of the EWs of absorption and emission in a given transition equals zero (Prochaska et al., in prep.).  The sum of the \ion{Mg}{2} EWs measured in the galaxy spectrum is $\sim 0.7$ \AA, indicating that the outflow is nearly isotropic, and hence that the sections of the wind observed in emission have $\tau_{2796} \gtrsim 1.4$ (i.e., similar to the portions of the wind observed along the line of sight).
With $v_\mathrm{wind} = 300 \mkms$, $\Delta v_\mathrm{wind} = 500 \mkms$, and $\Delta R_\mathrm{wind} = 7$ kpc,
and using the equation $\tau_{2796} = (\pi e^2 / m_e c) f_{2796} \lambda_{2796} n_{\mathrm{MgII}} (\Delta R_\mathrm{wind} / \Delta v_\mathrm{wind})$, modified from Eq.\ 8.45 of \citet{Lamers1999}, 
we find a number density $n_{\rm MgII} \sim 7 \times 10^{-9}~\rm cm^{-3}$.  Neglecting any ionization correction, and assuming solar abundance ($\log \rm Mg/H = -4.42$) and a factor of -1.2 dex dust depletion 
\citep{SavageSembach1996}, 
the corresponding limit on $n_H$ is $0.003~\rm cm^{-3}$.   
While this estimate suffers from large uncertainties, 
the limits on both the wind radius and $n_H$ determined from this analysis provide novel information which is crucial 
to constraining both the mass and energy carried by the outflow.

Beyond the analysis presented here, future studies of these emission features in other galaxies will provide critical insight into the impact of galactic winds on the redistribution metals and dust to the outer halos of galaxies \citep{Menard2009} and the IGM \citep{Simcoe2002}.  
A rest-frame UV spectroscopic survey of $0.3 < z < 1.4$ galaxies (Rubin et al.\ 2010, in prep.) reveals that these emission features are quite common at $z \sim 1$, and thus may be targeted for follow-up observations with IFUs to create maps of the outflow morphology.  High spectral resolution observations of \ion{Fe}{2}* emission will probe the 
gas kinematics in detail from the outskirts of the wind to deep within the galaxy, both on the far and near side of the stellar disk; in addition, the observed flux in the fine structure lines can provide an independent constraint on the radial extent of the gas \citep[e.g.,][]{Prochaska2006}.  These studies, when performed in concert with radiative transfer modeling of the emission features, will probe the morphology of outflows at an unprecedented level of detail in numerous distant galaxies.

\acknowledgements
The authors are grateful for support for this project from NSF grants AST-0808133 and AST-0507483.
J.X.P. acknowledges funding though an NSF CAREER grant (AST-0548180).
We thank R. da Silva and M. Fumagalli for providing code to coadd two-dimensional spectra.  



\clearpage
\begin{deluxetable}{lcccccccccc}
\tabletypesize{\footnotesize}
\tablecolumns{11}
\tablecaption{Observed Transitions and Limits \label{tab.em}}
\tablewidth{0pt}
\tablehead{\colhead{} & \colhead{$\rm E_{high}$} & \colhead{$\rm E_{low}$} & \colhead{J} & \colhead{$\lambda$} & \colhead{A} & \colhead{$W_r$} & \colhead{Flux\tablenotemark{a}} & \colhead{$v_{peak}$} & \colhead{$v_{cen}$} & \colhead{$v_{extent}$}\\
 & \colhead{$\rm cm^{-1}$} & \colhead{$\rm cm^{-1}$} &  & \colhead{\AA} & \colhead{$\rm s^{-1}$} & \colhead{\AA} & \colhead{$\rm 10^{-17} ergs~cm^{-2}~s^{-1}$} & \colhead{$\rm km~s^{-1}$} & \colhead{$\rm km~s^{-1}$} & \colhead{$\rm km~s^{-1}$}}
\startdata
     \ion{Fe}{2} UV1 & 38458.98 &     0.00 &    9/2$ \leftarrow $9/2 & 2600.17 & Absorption & $ 2.54 \pm  0.08$ & ... & ... & ... & ... \\
  & 38458.98 &   384.79 &   9/2$ \rightarrow $7/2 & 2626.45 & 3.41E+07 & $    -0.85 \pm   0.08$ & $0.959 \pm 0.082$ &    50 &   -28 &   -288$ \rightarrow $  218\\
                     & 38660.04 &     0.00 &    7/2$ \leftarrow $9/2 & 2586.65 & Absorption & $ 1.89 \pm  0.07$ & ... & ... & ... & ... \\
  & 38660.04 &   384.79 &   7/2$ \rightarrow $7/2 & 2612.65 & 1.23E+08 & $    -0.57 \pm   0.07$ & $0.756 \pm 0.092$ &   -62 &   -20 &   -317$ \rightarrow $  277\\
  & 38660.04 &   667.68 &   7/2$ \rightarrow $5/2 & 2632.11 & 6.21E+07 & $    -0.20 \pm   0.05$ & $0.266 \pm 0.064$ &   162 &   162 &     -6$ \rightarrow $  331\\
  & 38858.96 &   667.68 &   5/2$ \rightarrow $5/2 & 2618.40 & 4.91E+07 & ... & $ < 0.236$ & ... & ... & ...\\
  & 38858.96 &   862.62 &   5/2$ \rightarrow $3/2 & 2631.83 & 8.39E+07 & ... & b & ... & ... & ...\\
  & 39013.21 &   667.68 &   3/2$ \rightarrow $5/2 & 2607.87 & 1.74E+08 & ... & $ < 0.271$ & ... & ... & ...\\
  & 39013.21 &   862.61 &   3/2$ \rightarrow $3/2 & 2621.19 & 3.81E+06 & ... & $ < 0.251$ & ... & ... & ...\\
  & 39013.21 &   977.05 &   3/2$ \rightarrow $1/2 & 2629.08 & 8.35E+07 & ... & $ < 0.231$ & ... & ... & ...\\
  & 39109.31 &   862.61 &   1/2$ \rightarrow $3/2 & 2614.60 & 2.11E+08 & ... & $ < 0.267$ & ... & ... & ...\\
  & 39109.31 &   977.05 &   1/2$ \rightarrow $1/2 & 2622.45 & 5.43E+07 & ... & $ < 0.239$ & ... & ... & ...\\
\tableline \\ [-1.5ex]
     \ion{Fe}{2} UV2 & 41968.05 &     0.00 &   11/2$ \leftarrow $9/2 & 2382.76 & Absorption & $ 2.00 \pm  0.07$ & ... & ... & ... & ... \\
                     & 42114.82 &     0.00 &    9/2$ \leftarrow $9/2 & 2374.46 & Absorption & $ 1.37 \pm  0.08$ & ... & ... & ... & ... \\
  & 42114.82 &   384.79 &   9/2$ \rightarrow $7/2 & 2396.36 & 2.67E+08 & $    -0.81 \pm   0.08$ & $1.183 \pm 0.120$ &   119 &   -14 &   -430$ \rightarrow $  393\\
  & 42237.03 &     0.00 &   7/2$ \rightarrow $9/2 & 2367.59 & 3.21E+04 & ... & $ < 0.279$ & ... & ... & ...\\
  & 42237.03 &   384.79 &   7/2$ \rightarrow $7/2 & 2389.36 & 9.64E+07 & ... & $ < 0.301$ & ... & ... & ...\\
  & 42334.82 &   667.68 &   5/2$ \rightarrow $5/2 & 2399.97 & 1.37E+08 & ... & $ < 0.304$ & ... & ... & ...\\
  & 42401.30 &   667.68 &   3/2$ \rightarrow $5/2 & 2396.15 & 2.67E+07 & ... & b & ... & ... & ...\\
\tableline \\ [-1.5ex]
     \ion{Fe}{2} UV3 & 42658.22 &     0.00 &    7/2$ \leftarrow $9/2 & 2344.21 & Absorption & $ 2.08 \pm  0.08$ & ... & ... & ... & ... \\
  & 42658.22 &   384.79 &   7/2$ \rightarrow $7/2 & 2365.55 & 5.90E+07 & $    -0.17 \pm   0.06$ & $0.252 \pm 0.091$ &    45 &     0 &   -232$ \rightarrow $  230\\
  & 43238.59 &   384.79 &   5/2$ \rightarrow $7/2 & 2333.52 & 1.27E+08 & ... & $ < 0.301$ & ... & ... & ...\\
  & 43238.59 &   667.68 &   5/2$ \rightarrow $5/2 & 2349.02 & 1.09E+08 & ... & $ < 0.292$ & ... & ... & ...\\
  & 43238.59 &   862.62 &   5/2$ \rightarrow $3/2 & 2359.83 & 5.42E+07 & ... & $ < 0.284$ & ... & ... & ...\\
  & 43620.96 &   862.61 &   3/2$ \rightarrow $3/2 & 2338.73 & 1.09E+08 & ... & $ < 0.277$ & ... & ... & ...\\
\tableline \\ [-1.5ex]
    \ion{Mg}{2} & ... & ... & ... & 2796.35 & Absorption & $ 2.65 \pm  0.07$ & ... & ... & ... & ... \\
 & ... & ... & ... & ... & ... & $    -1.69 \pm   0.05$ & $  1.907 \pm 0.065$ &    60 &    89 &    -99$ \rightarrow $  297\\
 & ... & ... & ... & 2803.53 & Absorption & $ 1.62 \pm  0.05$ & ... & ... & ... & ... \\
 & ... & ... & ... & ... & ... & $    -1.84 \pm   0.06$ & $  2.086 \pm 0.079$ &   161 &   225 &    -76$ \rightarrow $  556\\
\tableline \\ [-1.5ex]
$\rm [$\ion{Ne}{5}$\rm ]$ & ... & ... & ... & 3426.98 & ... & $    -0.61 \pm   0.13$ & $  0.466 \pm 0.103$ &    51 &   212 &   -271$ \rightarrow $  695\\
 $\rm [$\ion{O}{2}$\rm ]$ & ... & ... & ... & 3727.10 & ... & $   -66.74 \pm   0.25$ & $ 50.263 \pm 0.190$ &   194 &   132 &   -464$ \rightarrow $  732\\
$\rm [$\ion{Ne}{3}$\rm ]$ & ... & ... & ... & 3869.84 & ... & $    -5.74 \pm   0.13$ & $  4.673 \pm 0.111$ &   -26 &    25 &   -373$ \rightarrow $  436\\
                H$\delta$ & ... & ... & ... & 4102.90 & ... & $    -4.84 \pm   0.15$ & $  3.591 \pm 0.117$ &   -39 &    38 &   -312$ \rightarrow $  397\\
                H$\gamma$ & ... & ... & ... & 4341.69 & ... & $   -14.42 \pm   0.30$ & $  7.912 \pm 0.207$ &     4 &    81 &   -407$ \rightarrow $  621\\
\enddata
\tablenotetext{a}{Line fluxes are determined by summing the continuum-subtracted spectrum over the velocity range of the line, as described in \S\ref{sec.mg2analysis}.}
\tablecomments{Rest-frame EW, flux and kinematic measurements from LRIS spectroscopy.  Upper and lower energy levels, orbital angular momenta (J) and Einstein coefficients are given for \ion{Fe}{2} transitions \citep{Morton2003}. 
Errors are $1\sigma$ uncertainties with $3\sigma$ limits.  Blended transitions are marked with ``b".  We check \ion{Fe}{2} identifications by comparing A coefficients among different transitions.  For optically thin gas, the flux ratio of two lines that originate from the same excited state is approximately the ratio of the corresponding A coefficients.  E.g., the flux ratio of the 2612 \AA \ and 2632.11 \AA \ transitions ($\rm E_{high} = 38660.04 \, cm^{-1}$)  is $\sim 2.8 \pm 0.7$, i.e.\ $\sim 1.2\sigma$ higher than the ratio of A coefficients (1.98).}
\end{deluxetable}

\clearpage
\end{document}